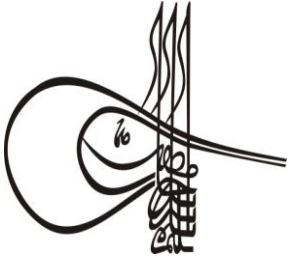

INTERNATIONAL
BALKAN
UNIVERSITY

EXCELLENCE FOR THE FUTURE
IBU.EDU.MK

# CHALLENGES IN INTEGRATING TECHNOLOGY INTO EDUCATION

*Oğuzhan ATABEK**

## ABSTRACT

Despite significant amount of investment, there seems to be obstacles to technology integration in education. In order to shed light on the nature of perceived obstacles to technology integration, opinions of 117 professionals, who were selected by Turkish Ministry of National Education as experts in their respective fields, about the obstacles to integration of technology into education were investigated. After categorizing the perceived obstacles by factor analysis, associations of those categories with personal and professional differences were further investigated for better contextualizing the findings. Correlations were analyzed by Pearson's product moment coefficient and point-biserial coefficient. The results revealed that it's not the hardware itself that constitute obstacles to technology integration. Insufficiency of in-service and pre-service training, content support, and incentive system emerged as major perceived obstacles to technology integration. Inadequacy of physical and technological infrastructure was also found to be an important obstacle to successful integration. Novelty of the technologies compared to older ones were not found to be an obstacle to technology integration. Moreover, participants stressed the lack of education in teacher training institutions about current technologies that Ministry of National Education officially requires teachers to use as part of their jobs to be another important obstacle. There were no correlations between sex, age, level of education, job position, year of experience in other careers, and any of the categories of perceived obstacles. However, there was a strong negative correlation between year of experience in teaching and

---

Dr., Akdeniz Üniversitesi Eğitim Fakültesi, E-posta: oguzhanatabek@gmail.com

insufficiency of resources. Association between year of experience in educational administration and negative psychological state was also strong and negative.

## STRUCTURED ABSTRACT

Despite significant amount of investment there seems to be obstacles to technology integration in education. In order to overcome the obstacles to successful, effective, and efficient implementation of educational technology, stakeholders should set sight on teachers. Teachers are tasked not only with introducing new technologies to learners but also with developing and delivering instruction that are designed with use of those technologies in mind. Moreover, as in Turkey, in most of the countries, governments are the most influential actors in education. A solution to the challenges technology integration is facing should not be designed without taking the government officials and administrators in national education establishment into account. In order to overcome obstacles, those obstacles should be identified. Identifying the problem is the most important step in problem analysis and decision making (Kepner & Tregoe, 2013; Lunenburg, 2010). Therefore, in order to contribute to the scientific understanding of integration of technology into education, a survey was conducted on professionals who were selected by Turkish Ministry of National Education (TMNE) on the basis of their expertise in their areas of responsibility and who attended the 19th National Education Council to identify the perceived obstacles to technology integration in Turkey. Associations of the categories of perceived obstacles with demographic variables were further investigated for exploring whether personal and professional differences can predict how obstacles to technology integration are perceived.

The study was designed as a quantitative research employing the environmental scanning method. Environmental scanning is a research method developed by Francis Joseph Aguilar and is used as part of strategic planning processes "in which emerging trends, changes and issues are regularly monitored and evaluated as to their likely impact" (Preble, Rau, & Reichel, 1988, p. 5). After the conclusion of 19th National Education Council, TMNE arranged a workshop on the goals and priorities of national education. From among the faculty members, ministerial and school-level administrators, and experienced teachers, TMNE selected and invited 147 experienced professionals as experts in their respective fields. After the conclusion of the workshop, all attendees were personally invited to participate in the survey. Out of 147 attendees, 117 participated in the study (N=117, 80%). Of the 117 participants, 18 were female (15.4%) and 99 were male (84.6%). Remarkably, 85 of participants (72.65%) had administrative duties during their carrier while the number of participants who are currently administrators was 64.

Data was collected by a paper-based survey consisting of a demographics questionnaire and an opinion questionnaire. Demographics questionnaire was for collecting basic demographics information, educational level, and information regarding respondents' career. Opinion questionnaire was developed by the researcher to elicit views of expert educators on obstacles to technology integration. It was based on the opinions of the attendees of the workshop who were selected by TMNE as experts on their fields. Attendees expressed a total of 21 issues as "obstacle to integration of technology into education". Researcher prepared a paper-based survey questionnaire consisting of 21 Likert-type 5-point multiple choice each representing one of those issues. Responses were "Strongly disagree", "Disagree", "Undecided", "Agree", and "Strongly agree". In addition to descriptive analysis on items of the questionnaires, a factor analysis was used to investigate whether perceived obstacles to technology integration fall into categories. It should be noted that, since the questionnaire was not intended for measuring psychological constructs, the factor analysis was not aimed for revealing latent variables representing psychological constructs. Rather, it was aimed for categorizing the items covariating in clusters. Overall, five categories were extracted from the questionnaire: Undersupply (U), Insufficiency of Resources (IR), Insufficiency of Infrastructure (INF), Negative Psychological State (NPS), Difficulty of Newer Technology (DNT). Finally, association between demographic variables and latent variables were analyzed by Pearson's product-moment correlation coefficient.

Analysis of the data revealed that means of 21 items ranged between 2.86 and 4.11. Item 19 yielded the highest mean of 4.11: "Lack of communication between educator, school, and universities". Item 20 was the one with the second greatest mean with the expression "Insufficiency of in-service training programs on effective use of information technologies". The lowest mean value (2.86) was obtained for Item 8: "Lack of some features of older technologies in newer ones". Remarkably, second lowest mean was calculated for Item 9 which was also questioning the difference between newer and older technologies: "New technologies are not as simple and easy to understand as older technologies".

As a category of perceived obstacles, U yielded the greatest mean (18.84). Items 15 and 17 were referring to lack of technological solutions that teachers can use. Item 19 which was the one with the greatest mean among all items was refereeing to lack of communication between educator, school, and universities. Item 20 was mentioning in-service training and finally, Item 21 was referring to content support for computers. IR was the category with the second greatest sample mean (17.96). INF was the intermediate category among five perceived obstacles group (10.61). NPS and DNT were the categories with the second lowest and lowest mean values, respectively. Apparently, educators do not see negative attitude towards, incompetence for, and difficulty in using technology as significant obstacles to technology integration.

Remarkably, there was no correlation between sex, age, level of education, job position, year of experience in non-educational careers and any of the categories of perceived obstacles. On the other hand, analyses with Pearson's product-moment correlation coefficient produced significant results for two of the demographic variables. There was a strong, negative correlation between year of experience in teaching and Insufficiency of Resources, which was statistically significant (r = -.254, n = 108, p = .008). Additionally, there was also a strong, negative correlation between year of experience in educational administration and Negative Psychological State, which was statistically significant (r = -.247, n = 84, p = .024). Both coefficients were negative signifying that perception of those two issues to be obstacles decreases as the year of experience in teaching or educational administration increases.

The purpose of the research was to investigate the opinions of professionals, who were selected by TMNE as experts in their respective fields, about the obstacles to integration of technology into education and to ascertain the associations between perceived obstacles and individual differences such as sex, age, level of education, and experience. Initial analysis revealed that experienced educators who work as teachers, school administrators, ministerial administrators, university faculty or education inspectors seem to unanimously think that hardware itself or novelty of it does not constitute an obstacle to technology integration. Participants' opinions revealed that, regarding technology integration, it's not the gadgets but the knowledge, information, and the processes which make the difference. Remarkably, supporting the research by Inan and Lowther (2010), there was no significant association between demographic variables and categories of perceived obstacles with the greatest, intermediate, and lowest mean values. Nonexistence of any relationship signifies that the perceptions of obstacles are stable irrespective of participants' sex, age, level of education, job position, and year of experience in teaching, administration or other careers.

In parallel with the findings of Fischer et al. (2018), in-service training was the most strongly agreed issue of which insufficiency leads to obstacles to technology integration. Results indicate that content to use with technology is more important to educators compared with the technology itself (Keser & Çetinkaya, 2013). It should be noted that, knowledge and skills that teachers demand from in-service training is also partly included in the curriculum of teacher training institutions. This study supported previous research (Jones & Madden, 2002; Mims, Polly, Shepherd, & Inan, 2006; O'Dwyer, Russell, & Bebel, 2004) reporting that "new graduates have more knowledge on technology integration … because teachers who recently graduated from a teacher preparation program would be more technology competent … and more prepared to integrate technology into classroom instruction" (Inan & Lowther, 2010, p.147) because of incorporation of technology into teacher training (Kirschner & Selinger, 2003). Education provided to pre-service teachers in those institutions may be improved to match the changing and growing needs of schools (Keser & Çetinkaya, 2013). Additionally, teacher training institutions may also provide education for the technologies that TMNE currently uses. Technologies currently in use such as e-School, EBA, MEBBIS, and DYS may be included in the curriculum of teacher training programs. Moreover, in parallel with the report of Stecher et al. (2018b), teachers seem to expect rewards for using time consuming technologies which demand further effort for learning and developing competence. Hence, an incentive system may help improve motivation for engagement in technology integration.

On the other hand, as argued by Dede (2011a, 2011b), sufficiency of technological and physical infrastructure is of vital importance for technology to be successfully integrated into education. As reported by Dede (2011b), teachers and schools should be provided with a habitat of technologies that flawlessly work together so that technology itself does not hinder instruction. Rather than a series of gadgets on the teacher's desk, technology should be conceptualized as a "solution" to the system of instruction. An effectively working system comprised of infrastructure, content, pre-service and in-service training, incentives, and harmony between technology and curriculum (Okita & Jamalian, 2011; Topuz & Kaptan, 2017) may bring investment in educational technology to a successful conclusion. As experience in educational administration increases, perception of negative attitude and incompetence as obstacles to technology integration decreases. This finding indicates that negative attitude and incompetence may be overcome by a well-designed system of educational technology solution combined with support of teachers by training, content and incentives. Hence, a habitat of knowledge, information, and the processes are key to succeeding with educational technology.

**Keywords:** Educational Technology, Technology Integration, Teacher Training, In-service Training, Educational Content.

## TEKNOLOJİNİN EĞİTİMLE BÜTÜNLEŞTİRİLMESİNDE KARŞILAŞILAN GÜÇLÜKLER

### ÖZET

Eğitim teknolojisi alanında yapılan önemli miktarda yatırımlara karşın, teknolojinin eğitim ile bütünleştirilmesinin önünde hala önemli engeller bulunmaktadır. Bu araştırmada, Milli Eğitim Bakanlığı tarafından kendi alanlarında uzman olarak seçilen 117 profesyonelin, teknolojinin eğitimle bütünleştirilmesine yönelik algıladıkları engellerin ortaya çıkarılması amaçlanmıştır. Ayrıca araştırmada, bulguların daha iyi bağlamsallaştırılabilmesi için faktör analizi kullanılarak algılanan engeller kategorileştirildikten sonra işbu algılanan engel kategorileri ile kişisel ve profesyonel farklılıklar arasındaki ilişkiler de incelenmiştir. İlişkiler; Pearson momentler çarpımı korelasyon katsayısı ve nokta çift serili korelasyon katsayısı ile çözümlenmiştir. Elde edilen bulgular, hizmet içi ve hizmet öncesi eğitim, içerik desteği ve teşvik sistemindeki yetersizliklerin teknolojinin bütünleştirilmesine yönelik başlıca engeller olarak algılandığını göstermiştir. Ayrıca, katılımcıların görüşlerine göre, fiziksel ve teknolojik altyapının yetersizliğinin de başarılı bir bütünleştirmeye yönelik önemli bir engel olduğu sonucuna varılmıştır. Ek olarak, katılımcılar, Milli Eğitim Bakanlığının mesleklerinin bir parçası olarak öğretmenlerden kullanmalarını resmen istediği güncel teknolojilerin eğitiminin öğretmen yetiştiren kurumlarda verilmesini de diğer bir önemli engel olarak vurgulamışlardır. Cinsiyet, yaş, eğitim düzeyi, mesleki pozisyon ve eğitim dışı diğer kariyerlerdeki yıl türünden deneyim ile algılanan engellerin kategorileri arasında anlamlı bir ilişki bulunmamıştır. Ancak; öğretmenlikteki yıl türünden deneyim ile olanakların yetersizliği kategorisi arasında güçlü ve negatif bir ilişki bulunmuştur. Eğitim yöneticiliğindeki yıl türünden deneyim ile olumsuz psikolojik durum kategorisi arasında da yine güçlü ve negatif bir ilişki bulunmuştur.

**Anahtar Kelimeler:** Eğitim Teknolojisi, Teknoloji Entegrasyonu, Öğretmen Yetiştirme, Hizmetiçi Eğitim, Eğitimsel İçerik.

### Introduction

Technology was once used to be viewed as redundant tools for education. Even technologies like blackboard and book were viewed as radical tools when they were first introduced (Haran, 2015). Educational technology is now considered to be inseparable from learning (Jegede, Fraser, & Curtin, 1995), development and research in science (Guzey & Roehrig, 2012), and even the very notion of technology (Putri, 2016). However, educational technology "has not been fully integrated into the field of education" (Guzey & Roehrig, 2012, p. 62). On the other hand, tremendous amount of financial resources has been invested in educational technology. Adkins (2018) reports that, in 2017, global funding going to global educational technology companies reached a new record of $9.56 billion. Between 1997 and 2017, $37.8 billion were invested in educational technology companies and 62% of that amount was invested in the last three years (Shulman, 2018). Yet, most of the teachers struggle to use (Rebora, 2016) or do not use technology in meaningful ways (Guzey & Roehrig, 2012). Moreover, misuse of technology is becoming widespread in the classrooms (Fox, 2018, p. 28; Glendinning, 2018; Hyndman, 2018; Ribble & Bailey, 2004). Hence, in spite of considerable amount of investment, there seems to be obstacles to technology integration in education.

Another issue with technology integration is the impact of technology on how teachers actually teach. Even when it is implemented, educational technology seems to be not transforming how teachers teach. Herold (2015) argues that teachers have been significantly slow to transform the ways they teach. Bill & Melinda Gates Foundation designed and funded "The Intensive Partnerships for Effective Teaching" initiative, launching in 2009, as a multiyear effort to dramatically improve student outcomes by increasing students' access to effective teaching (Gutierrez, Weinberger, & Engberg, 2018). RAND Corporation conducted a six-year evaluation of the program and concluded that the program failed to achieve its goals for improved student achievement and graduation (Stecher et al., 2018a, p. 1). RAND Corporation noted that making lasting changes to how teachers teach is difficult and the results from the evaluation of the initiative demonstrate the challenges of getting schools and teachers to embrace big changes (Will, 2018). Remarkably, teachers were highly motivated to change their teaching for lessons that were going to be observed in order "to improve their observation scores and receive salary increases", however, "they were not so motivated to make lasting changes in their practice" (Stecher et al., 2018b, pp. 280-281). Incentives for encouraging and motivating teachers to integrate educational technology into their lessons seem to be more crucial than what is believed to be especially if the goal is to make lasting changes in teachers' practice.

In order to put the importance of factors influencing technology integration into perspective, significance of educational technology should be taken into consideration. Oxford Dictionaries defines technology as "the application of scientific knowledge for practical purposes, especially in industry" ("Technology", 2018b). Merriam-Webster Dictionaries defines it as "the practical application of knowledge especially in a particular area" and "a capability given by the practical application of knowledge" ("Technology", 2018a). Borgmann (2006) argues that technology forms and changes the culture and is harmful when used injudiciously or excessively. Therefore, educational technology is the practical application of knowledge for educational purposes. Association for Educational Communications and Technology defines educational technology as "the study and ethical practice of facilitating learning and improving performance by creating, using and managing appropriate technological processes and resources" (Januszewski & Molenda, 2008, p. 1). Remarkably, definition of the field is unsparing in the use of the word "ethical". Emphasis on ethics evokes Borgmann's argument regarding injudicious and excessive use of technology (2006) and Fox's argument regarding misuse of technology (2018).

It is not hard to estimate the vitality of technology in education especially considering the importance, role, and impact of technology in our daily lives. Educational technology is an extensive part of modern education and may give way to far-reaching undesired consequences when it is not properly employed. For example, contemporary distance education is almost entirely provided by Internet based educational technologies. According to Distance Education Enrollment Report of Babson Survey Research Group, as of 2017, 29.7% of all students are taking at least one distance course in higher education and 83.0% of those students are studying at the undergraduate level (Allen & Seaman, 2017, p. 4). Additionally, Technology is argued to be one of the most effective factors in school improvement "not only for the purpose of teaching and learning but also for administrative use" (Ghavifekr, Afshari, Siraj, & Seger, 2013, p. 1344). Educational institutions with thousands of students almost completely rely on technology for the management of the institution and the service they provide. Therefore, integration of technology into education is vital, "inevitable and cannot be avoided" (Ghavifekr, Afshari, Siraj, & Seger, 2013, p. 1344). Hence, obstacles to integration should be investigated, determined, and countered with practical solutions in order to achieve successful, efficient, and productive implementation of educational technology.

According to "Education at a glance 2018" report published by Organisation for Economic Co-operation and Development, Turkey is not one of the leading countries among OECD countries regarding education (OECD, 2018). According to Turan (2002), technology and its use has not been institutionalized in educational establishment including public universities in Turkey. He argues that Turkey's fundamental educational problems ranging from general understanding of education to current situation of classrooms hinder effective integration of technology into education. Nevertheless, in 2011, Turkish government put an ambitious plan into effect for enhancing technology integration, titled "FATIH Project". The project was defined as "the greatest and most comprehensive educational movement regarding the use of technology in education ever carried into effect in the world, which is designed for every student to reach the best education, top-quality content, and for establishing the equality of opportunity in education" ("About FATIH Project", n.d.). Project aimed at broadband internet and multifunction printer for every school; interactive whiteboard and wireless network for every classroom; tablet computer, access to a government sponsored educational content website, e-mail account, cloud account, learning management system account, and content development studio for every teacher; and finally a digital ID, tablet computer, access to a government sponsored educational content website, mail account, cloud account, and access to personalized instructional material for every student. However, after 7 years of large-scale investment, the project remained inconclusive and Turkish Ministry of National Education (TMNE) shifted its focus to "process based and instrumentalist" approaches which take "functionality" into consideration and which "concentrate upon the content" (Ministry of National Education, 2018). Turkey's bitter experience with large-scale investment in educational technology indicates that integration of technology into education is a major problem in Turkey too.

In order to overcome the obstacles to successful, effective, and efficient implementation of educational technology, stakeholders should set sight on teachers. Teachers are tasked not only with introducing new technologies to learners but also with developing and delivering instruction that are designed with use of those technologies in mind. Stecher et al. state that "teachers remain the most salient in-school factor in determining student outcomes, and thus improving teaching is a plausible lever for improvement" (2018b, p. 502). But getting teachers embrace new changes to their way of teaching is difficult "especially when state and local policies are in flux" (Will, 2018). Moreover, changes in governments' laws, regulations, and practices influence the implementation of the reforms (Stecher et al., 2018a, p. 7). As in Turkey, in most of the countries, governments are the most influential actors in education. For instance, "public institutions continue to educate the largest proportion of distance students" (Allen & Seaman, 2017, p. 4). A solution to the challenges technology integration is facing should not be designed without taking the government officials and administrators in national education establishment into account. Therefore, administrators working in ministry of education in addition to teachers and faculty members may be considered as more influential actors on the issue of technology integration.

In order to overcome obstacles, those obstacles should be identified. Identifying the problem is the most important step in problem analysis and decision making (Kepner & Tregoe, 2013; Lunenburg, 2010). Therefore, in order to contribute to the scientific understanding of integration of technology into education, a survey was conducted on professionals who were selected by TMNE on the basis of their expertise in their areas of responsibility and who attended the 19[th] National Education Council to identify the perceived obstacles to technology integration in Turkey. Associations of the categories of perceived obstacles with demographic variables were further investigated for exploring whether personal and professional differences can predict how obstacles to technology integration are perceived.

**Method**

The study was designed as a quantitative research employing the environmental scanning method. Environmental scanning is a research method developed by Francis Joseph Aguilar and is used as part of strategic planning processes "in which emerging trends, changes and issues are regularly monitored and evaluated as to their likely impact" (Preble, Rau, & Reichel, 1988, p. 5). It can be used to identify important emerging issues that may constitute either obstacles or opportunities (Masini, 1993; Renfro & Morrison, 1984). Environmental scanning enables decision-makers to understand current and potential changes taking place in their institution's external environment (Fahey, King, & Narayanan, 1981). Since the purpose of the study was to identify the obstacles to integration of technology into education and was to evaluate the obstacles on the basis of their likely impact, environmental scanning was chosen as the research method of the study. Environmental scanning may contribute to organizational learning and organizational ability to deal with rapid changes that are taking place (Jain, 1984; McEwen, 2008; Voros, 2003). By choosing environmental scanning method, it was aimed to produce scientific knowledge which Turkish national education establishment may use for adapting to technological developments that are currently taking place. All procedures were in accordance with the APA Ethical Principles of Psychologists and Code of Conduct.

**Study group**

After the conclusion of 19[th] National Education Council, TMNE arranged a workshop on the goals and priorities of national education. From among the faculty members, ministerial and school-level administrators, and experienced teachers, TMNE selected and invited 147 experienced professionals as experts in their respective fields. After the conclusion of the workshop, all attendees were personally invited to participate in the survey. Out of 147 attendees, 117 participated in the study (N=117, 80%). Of the 117 participants, 18 were female (15.4%) and 99 were male (84.6%). Overrepresentation of males in administrative positions caused the gender gap in the sample. Ages of the respondents ranged between 26 and 63 with a mean of 44.08 (M = 44.08, SD = 8.42). Among the participants, eight had PhD (6.8%), 42 had MSc (35.9%), and 67 had BSc (57.3%) degrees. Job positions of the participants are depicted in Table 1.

**Table 1. Job positions of the participants.**

| Position | f | % |
|---|---|---|
| Teacher without an administrative duty | 37 | 31.6 |
| Educational administrator at a school | 56 | 47.9 |
| Educational administrator at provincial government post | 5 | 4.3 |
| Educational administrator at central government post | 3 | 2.6 |
| Specialist | 6 | 5.1 |
| Inspector | 5 | 4.3 |
| Faculty member | 5 | 4.3 |
| Total | 117 | 100 |

*Note: f represents frequency and % represents percentage.*

Of all the participants, 6 were specialists and hence not teachers. Remaining 111 individuals were teachers. Almost half of the participants (47.9%) were educational administrator at a school. Experience of participants in terms of years is depicted in Table 2. Remarkably, 85 of participants (72.65%) had administrative duties during their carrier while the number of participants who are currently administrators was 64. Participants were relatively experienced both in teaching and administration in terms of years they worked as administrators. Participants had up to 36 years in teaching and up to 34 years in administration.

**Table 2. Experience of participants in years.**

| Position | f | % | Minimum | Maximum | Mean | SD |
|---|---|---|---|---|---|---|
| In teaching | 111 | 94.87 | 1 | 36 | 15.09 | 8.40 |
| In administration | 85 | 72.65 | 1 | 34 | 11.34 | 8.24 |
| In other positions | 23 | 19.65 | 1 | 31 | 8.39 | 8.70 |

*Note: f represents frequency, % represents percentage, and SD stands for standard deviation.*

### Data collection

Environmental scanning is defined as systematic collection of information for lessening the randomness of information and for providing early warnings for managers of changing external conditions (Aguilar, 1967). While conducting environmental scanning, researchers seek for information about signals of change in the social, technological, economic, environmental, and political categories (Du Toit, 2016; Morrison, 1993). In this study, information about signals of change was sought for in the technological category. Information may be obtained from scientific and non-scientific publications, TV and radio programs, conferences, and from knowledgeable individuals in researcher's personal information network (Morrison, 1993). In this research, information was obtained from knowledgeable individuals who were attending a workshop on the goals and priorities of national education and who were in researcher's information network through the workshop.

Data was collected by a paper-based survey consisting of a demographics questionnaire and an opinion questionnaire. Demographics questionnaire was for collecting basic demographics information, educational level, and information regarding respondents' career. Opinion questionnaire was developed by the researcher to elicit views of expert educators on obstacles to technology integration. It should be noted that, the opinion questionnaire is not a psychometric scale for measuring psychological constructs.

### Development of opinions questionnaire

Opinion questionnaires are "one of the most useful research instruments in social psychology" (Politz, 1953, p. 11). Koch (2018) states that "a survey's questions must reflect a population's interests and opinions" (p. 2) for reflecting the sentiments of respondents and "opinion questionnaire is only as good as the representative sample of the population that responds to a request for participation" (p. 4). Hence, questionnaire was based on the opinions of the attendees of the workshop who were selected by TMNE as experts on their fields. In the course of the workshop, during the commission meetings and personal communications, the researcher noted the issues that were expressed by workshop attendees to be obstacles to technology integration. Attendees expressed a total of 21 issues as "obstacle to integration of technology into education". Researcher prepared a paper-based survey questionnaire consisting of 21 items each representing one of those issues.

Questionnaire included 21 Likert-type 5-point multiple choice items. Responses were "Strongly disagree", "Disagree", "Undecided", "Agree", and "Strongly agree". Questionnaire was designed to measure the degree to which individuals believe that particular issues are obstacles to integration of technology into education. Questionnaire included items such as: "Insufficiency of physical infrastructure of educational institutions", "New technologies are not as simple and easy to understand as older technologies", "Developing materials by using information technologies takes too much time", and "Insufficiency of in-service training programs on effective use of information technologies".

A factor analysis was used to investigate whether perceived obstacles to technology integration fall into categories. It should be noted that, since the questionnaire was not intended for measuring psychological constructs, the factor analysis was not aimed for revealing latent variables representing psychological constructs. Rather, it was aimed for categorizing the items covariating in clusters. Overall, five categories were extracted from the questionnaire: Undersupply (U, 5 items), Insufficiency of Resources (IR, 5 items), Insufficiency of Infrastructure (INF, 3 items), Negative Psychological State (NPS, 3 items), Difficulty of Newer Technology (DNT, 3 items). Statistical results of factor analysis are depicted in Table 3.

*Table 3. Categories and related statistical information.*

| Factor | Min. | Max. | $\bar{x}$ | s | Factor Loading | Variance Explained | Cronbach's α |
|---|---|---|---|---|---|---|---|
| Undersupply | 9 | 25 | 18.84 | 3.13 | 0.64 - 0.49 | 10.85 | 0.65 |
| Insufficiency of Resources | 6 | 25 | 17.96 | 3.49 | 0.71 - 0.49 | 5.81 | 0.66 |
| Insufficiency of Infrastructure | 3 | 15 | 10.61 | 2.74 | 0.91 - 0.56 | 9.04 | 0.81 |
| Negative Psychological State | 3 | 15 | 9.97 | 2.53 | 0.75 - 0.65 | 8.01 | 0.68 |
| Difficulty of Newer Technology | 3 | 15 | 8.97 | 2.73 | 0.82 - 0.62 | 24.12 | 0.71 |

**Note:** *Min., Max., $\bar{x}$, and s stand for minimum, maximum, sample mean, and sample standard deviation, respectively.*

**Procedure**

Workshop on the goals and priorities of national education was arranged as ten commissions. The researcher was the coordinator of the commission titled "Increasing the prevalence of the use of technological tools in education". Before the ending of workshop, researcher got permission from the ministry for conducting the research. After the conclusion of the workshop, attendees were invited to participate in the study by completing the survey. The researcher obtained informed consent from attendees for participating in the research and they were informed about the confidentiality of the information they provided. It is argued that opinions of structured groups such as an expert group are more accurate than those unstructured groups consisting of non-experts (Rowe & Wright, 2001). Hence, collecting opinions from expert groups contributes to the validity and reliability of the research.

**Data analysis**

Initially, all papers were screened to exclude those which were not completed. Then, answers of the respondents were transferred to computer. Data was stored, arranged, reviewed and analyzed by the use of IBM SPSS Statistics computer program (IBM SPSS Statistics Version 22). Demographic information and items were analyzed by descriptive statistical measures. However, there are no standard procedures for analyzing the information collected in an environmental scanning research (Du Toit, 2016; Sewdass & Du Toit, 2014). In addition to descriptive analysis on items of the questionnaires developed through first stages, items may be categorized by statistical factor analysis (Hambrick, 1982; West, & Anthony, 1990) for clustering items into categories. Factor analysis technique "serves several related purposes" (DeVellis, 2003, p. 103). In addition to determining psychological constructs, factor analysis can provide a means for condensing information by explaining variation among many items by using newly created variables (DeVellis, 2016). This second purpose of factor analysis is intended for enabling "variation to be accounted for by using a smaller number of variables" (DeVellis, 2016, p.

117). In this arrangement, factor analysis is employed as a "variable reduction procedure" (O'Rourke, Psych, & Hatcher, 2013, p. 2).

In this research study, rather than determining psychological constructs, factor analysis was employed for clustering covariating items into categories of perceived obstacles so that variation can be accounted for by using a smaller number of variables. Principal component analysis with varimax rotation was employed in order to investigate whether items were clustering into categories. Emerging factors (categories) were calculated into latent variables by summation of the scores. Finally, association between demographic variables and latent variables were analyzed by Pearson's product-moment correlation coefficient and point-biserial correlation coefficient.

**Results**

Data was analyzed in three phases. First, descriptive statistics was used for demographics and items of the opinion questionnaire. Then, a principal component analysis with varimax rotation was carried out to investigate the factor structures. Reliability analysis was also conducted. Finally, inferential statistics was used to investigate the associations between demographic variables and factors extracted in the second phase.

**Descriptive analysis of the questionnaire**

Items were answered by one of the choices from among "strongly disagree", "disagree", "undecided", "agree" and "strongly agree". Choices numerically corresponded to 1, 2, 3, 4, and 5, respectively. Means of the items were calculated by summation of respective scores of each case. Means could theoretically have a minimum value of 1 and a maximum value of 5. Analysis of the data revealed that means of 21 items ranged between 2.86 and 4.11. Means, standard deviations, and variances of items are depicted in Table 4.

**Table 4. Items and related descriptive results in descending order of means.**

| Items | $\bar{x}$ | s | $s^2$ |
|---|---|---|---|
| Q19 Lack of communication between educator, school, and universities (U) | 4.11 | 0.83 | 0.69 |
| Q20 Insufficiency of in-service training programs on effective use of information technologies (U) | 3.87 | 1.00 | 1.00 |
| Q14 Lack of reward for educators' use of information technologies by an incentive system (IR) | 3.83 | 0.95 | 0.91 |
| Q21 Lack of content support for certain technologies like tablet computer (U) | 3.79 | 0.98 | 0.96 |
| Q03 Lack of sufficient education in teacher training institutions about effective use of information technologies (IR) | 3.75 | 1.02 | 1.05 |
| Q04 Lack of sufficient education in teacher training institutions about learning management environments like "e-School" (IR) | 3.63 | 1.12 | 1.25 |
| Q16 Lack of sufficient evaluation of students' use of information technology | 3.63 | 0.85 | 0.73 |
| Q05 Insufficiency of physical infrastructure of educational institutions (INF) | 3.62 | 1.13 | 1.27 |
| Q10 Compared with older technologies, new ones require constantly learning | 3.61 | 1.10 | 1.21 |
| Q01 Incompetence of educators regarding the effective use of information technologies for educational purposes (NPS) | 3.57 | 1.03 | 1.06 |
| Q15 Lack of sufficient information technology solutions that educators can use for evaluation and assessment (U) | 3.56 | 1.05 | 1.11 |
| Q06 Insufficiency of technological infrastructure of educational institutions (INF) | 3.54 | 1.12 | 1.26 |
| Q17 Lack of information technology solutions that are sensitive to individual differences of students (U) | 3.52 | 0.94 | 0.88 |
| Q07 Lack of freely available content which is appropriate for effective use for educational purposes (INF) | 3.44 | 0.97 | 0.94 |
| Q18 Corporate culture in educational institutions conduces to resistance to new technologies (NPS) | 3.42 | 1.07 | 1.14 |
| Q13 Developing materials by using information technologies takes too much time (IR) | 3.38 | 1.07 | 1.15 |
| Q12 Educators inability to allocate time for use of information technologies because of their work load (IR) | 3.34 | 1.17 | 1.37 |
| Q11 Inappropriateness of the curriculum for effective use of information technologies (DNT) | 3.21 | 1.17 | 1.38 |
| Q02 Negative attitude of educators towards information technologies (NPS) | 2.97 | 1.13 | 1.27 |
| Q09 New technologies are not as simple and easy to understand as older technologies (DNT) | 2.86 | 1.16 | 1.35 |
| Q08 Lack of some features of older technologies in newer ones (DNT) | 2.86 | 1.08 | 1.17 |

*Note: $\bar{x}$, s, and $s^2$ stands for sample mean, standard deviation, and variance, respectively.*

Item 19 yielded the highest mean of 4.11. Item 19 was questioning communication and support by the expression "Lack of communication between educator, school, and universities". Item 19 refers to training of teachers by -faculties of- universities. Therefore, it is evident that educators' strong positive response for Item 19 indicates educators' need and desire for up-to-date scientific knowledge and related skills that "communication with university" may provide. Experienced educators' emphasis on communication and support reveals (a) their cognizance in the importance of technology for education, (b) their needs regarding the integration of technology, and (c) how to meet those needs.

Item 20 was the one with the second greatest mean with the expression "Insufficiency of in-service training programs on effective use of information technologies". Yielding a mean value of 3.87, the item was questioning the perceived impact of in-service training programs in technology integration.

A considerable proportion of participants seems to believe that insufficiency of such programs are an obstacle to technology integration. Remarkably, first two items with the greatest mean values were about teachers' need for knowledge and skills that a higher authority such as universities may provide. Teachers are aware that competence in technology integration requires specific knowledge and skills. Moreover, those knowledge and skills are strongly believed to be considerably sophisticated and demanding in a way that acquiring them necessitates a higher authority like universities or institutions providing in-service training.

Third greatest mean value was yielded by Item 14: "Lack of reward for educators' use of information technologies by an incentive system". Compared to first two items with greatest and second greatest means, Item 14 was questioning teachers' opinion about incentives. Mean of Item 14 was third only with a 0.04 difference from second greatest. Therefore, lack of in-service training for access to knowledge and skills, and lack of incentives for implementation of them seem to be somewhat leading perceived obstacles to technology integration.

The lowest mean value (2.86) was obtained for Item 8: "Lack of some features of older technologies in newer ones". Remarkably, second lowest mean was calculated for Item 9 which was also questioning the difference between newer and older technologies: "New technologies are not as simple and easy to understand as older technologies". These results indicate that experienced educators do not believe that newness or novelty of technologies constitute an obstacle to technology integration. Item 2 was the third lowest one: "Negative attitude of educators towards information technologies". Items 2, 8, and 9 were the only ones which had means lower than 3 (Undecided).

**Categories of perceived obstacles**

Principal component analysis with varimax rotation was used to investigate categories of perceived obstacles. It should be noted that the factor analysis was not intended for producing latent variables for psychological constructs. Rather, the purpose was to investigate if perceived obstacles fall into categories. Categorizing perceived obstacles into groups was intended for facilitating the process of describing the current situation. Undersupply (U), Insufficiency of Resources (IR), Insufficiency of Infrastructure (INF), Negative Psychological State (NPS), Difficulty of Newer Technology (DNT) emerged as categories of perceived obstacles. Table 4 depicts the factors (categories) extracted from the questionnaire.

As a category of perceived obstacles, U yielded the greatest mean (18.84). Items 15 and 17 were referring to lack of technological solutions that teachers can use. Item 19 which was the one with the greatest mean among all items was refereeing to lack of communication between educator, school, and universities. Item 20 was mentioning in-service training and finally, Item 21 was referring to content support for computers. Remarkably, expert educators refrained from conceptualizing obstacles to technology integration as a matter of hardware. Even when it was about computers, teachers choose to address the issue as a "solution" or "content". Solution is a broader term indicating a habitat of information technologies rather than just piece of hardware. On the other hand, content is the material that strategies, methods, and techniques of instruction are delivered through a medium. Perceived obstacles in U category –as opposed to the mere installation of hardware- emphasizes the knowledge and information that is required for properly employing technology for educational purposes.

IR was the category with the second greatest sample mean (17.96). Items 3 and 4 were referring to the required technology education provided in teacher training institutions. Items 12 and 13 were mentioning the time needed for using and developing materials with information technologies. Knowledge acquired through education and time were grouped as "resource". Item 14 was referring to reward for educators by an incentive system for using information technologies. According to mean values, Item 14 was the third greatest one among all items, indicating the excessive emphasis put on incentives by teachers.

INF was the intermediate category among five perceived obstacles group (10.61). Items 5 and 6 were referring to insufficiency of the physical and technological infrastructure of educational institutions. Item 7 was mentioning the lack of content that are freely accessible by teachers for using with the information technologies. Apparently, expert educators conceive technology integration in a two-fold manner: (1) physical and technological infrastructure and (2) content to use on or with that infrastructure. Results regarding IR and INF endorsed the findings of Burak, Özmenteş, & Seban (2015) about inadequacy of time, material, and physical conditions of the classroom.

NPS and DNT were the categories with the second lowest and lowest mean values, respectively. Items 2 and 18 were referring to the negative attitude of educators towards information technologies while Item 1 was referring to the incompetence of educators regarding those technologies. Items 8 and 9 were mentioning the differences between older and newer technologies with respectively lowest and second lowest mean values among all items. Item 11 was referring to the mismatch between curriculum and use of information technologies. Apparently, educators do not see negative attitude towards, incompetence for, and difficulty in using technology as obstacles to technology integration.

### Results from correlational analysis

Even though purpose of the factor analysis was not producing variables for psychological constructs, extracted factors which represent categories of perceived obstacles to technology integration were further analyzed by correlational statistical techniques to contextualize the findings from descriptive and factor analyses. Remarkably, there were no correlations between sex, age, level of education, job position, year of experience in non-educational careers and any of the categories of perceived obstacles. On the other hand, analyses with Pearson's product-moment correlation coefficient produced significant results for two of the demographic variables. There was a strong, negative correlation between year of experience in teaching and Insufficiency of Resources, which was statistically significant ($r = -.254$, $n = 108$, $p = .008$). Additionally, there was also a strong, negative correlation between year of experience in educational administration and Negative Psychological State, which was statistically significant ($r = -.247$, $n = 84$, $p = .024$). Both coefficients were negative signifying that perception of those two issues to be obstacles decreases as the year of experience in teaching or educational administration increases.

It should be noted that Insufficiency of Resources decreased only with an increase in year of experience in teaching. Similarly, Negative Psychological State decreased only with an increase in year of experience in educational administration. These two categories were not correlated with other demographic variables: sex, age, level of education, job position, and year of experience in non-educational careers. Moreover, Undersupply did not correlate with any of the demographic variables and stands firm as the leading category of obstacles. Insufficiency of Infrastructure is the intermediate-mean category and Difficulty of Newer Technology is the lowest-mean category irrespective of sex, age, level of education, job position, year of experience in non-educational careers, year of experience in teaching, and year of experience in educational administration.

### Discussion

The purpose of the research was to investigate the opinions of professionals, who were selected by TMNE as experts in their respective fields, about the obstacles to integration of technology into education and to ascertain the associations between perceived obstacles and individual differences such as sex, age, level of education, and experience. Initial analysis revealed that experienced educators who work as teachers, school administrators, ministerial administrators, university faculty or education inspectors seem to unanimously think that hardware itself or novelty of it does not constitute an obstacle to technology integration. Participants' opinions revealed that, regarding technology integration, it's not the gadgets but the knowledge, information, and the processes which make the difference. Remarkably, supporting the research by Inan and Lowther (2010), there was no significant association between demographic variables and categories of perceived obstacles with the greatest, intermediate, and lowest mean values. Nonexistence of any relationship signifies that the perceptions of obstacles are stable irrespective of participants' sex, age, level of education, job position, and year of experience in teaching, administration or other careers.

In parallel with the findings of Fischer et al. (2018), in-service training was the most strongly agreed issue of which insufficiency leads to obstacles to technology integration. Results indicate that content to use with technology is more important to educators compared with the technology itself (Keser & Çetinkaya, 2013). It should be noted that, knowledge and skills that teachers demand from in-service training is also partly included in the curriculum of teacher training institutions. This study supported previous research (Jones & Madden, 2002; Mims, Polly, Shepherd, & Inan, 2006; O'Dwyer, Russell, & Bebel, 2004) reporting that "new graduates have more knowledge on technology integration … because teachers who recently graduated from a teacher preparation program would be more technology competent … and more prepared to integrate technology into classroom instruction" (Inan & Lowther, 2010, p.147) because of incorporation of technology into teacher training (Kirschner & Selinger, 2003). Education provided to pre-service teachers in those institutions may be improved to match the changing and growing needs of schools (Keser & Çetinkaya, 2013). Additionally, teacher training institutions may also provide education for the technologies that TMNE currently uses. Technologies currently in use such as e-School, EBA, MEBBIS, and DYS may be included in the curriculum of teacher training programs. Moreover, in parallel with the report of Stecher et al. (2018b), teachers seem to expect rewards for using time consuming technologies which demand further effort for learning to use and developing competence for them. Hence, an incentive system may help improve motivation for engagement in technology integration.

On the other hand, as argued by Dede (2011a, 2011b), sufficiency of technological and physical infrastructure is of vital importance for technology to be successfully integrated into education. As reported by Dede (2011b), teachers and schools should be provided with a habitat of technologies that flawlessly work together so that technology itself does not hinder instruction. Rather than a series of gadgets on the teacher's desk, technology should be conceptualized as a "solution" to the system of instruction. An effectively working system comprised of infrastructure, content, pre-service and in-service training, incentives, and harmony between technology and curriculum (Okita & Jamalian, 2011; Topuz & Kaptan, 2017) may bring investment in educational technology to a successful conclusion. As experience in educational administration increases, perception of negative attitude and incompetence as obstacles to technology integration decreases. This finding indicates that negative attitude and incompetence may be overcome by a well-designed system of educational technology solution combined with support of teachers by training, content and incentives. Hence, a habitat of knowledge, information, and the processes are key to succeeding with educational technology.